\documentclass{article}

\usepackage[right= 0.75in, left= 0.75in, top=1.05in, bottom=1.15in]{geometry}

\usepackage{graphicx}
\usepackage[misc]{ifsym}
\usepackage{amsmath}
\usepackage{amsfonts}
\usepackage{mathrsfs} 
\usepackage{booktabs} 
\usepackage{array} 
\usepackage{paralist} 
\usepackage{verbatim} 
\usepackage{subfig} 
\usepackage{empheq}
\usepackage{bm}
\usepackage{tabularx}
\usepackage{stmaryrd}
\usepackage[affil-it,auth-sc]{authblk}
\usepackage[colorlinks, citecolor=cyan, linkcolor=cyan]{hyperref}

\DeclareMathOperator{\divg}{\mbox{div}} 
\DeclareMathOperator{\grad}{\mbox{grad}} 

\newcommand{\mb}[1]{\mathbf{#1}}

\newcommand{\mc}[1]{\mathcal{#1}} 

\author[1]{Alessandro Lucantonio$^*$}
\author[1]{Giovanni Noselli\thanks{Authors contributed equally.}\thanks{Corresponding author -- e-mail address:\,\texttt{giovanni.noselli@sissa.it}.}}
\affil[1]{\small{SISSA--International School for Advanced Studies, via Bonomea 265, 34136 Trieste, Italy.}}

\title{Concurrent factors determine toughening in the hydraulic fracture of poroelastic composites}

\date{}

\begin{document}

\maketitle

\begin{abstract}
Brittle materials fail catastrophically. In consequence of their limited flaw-tolerance, failure occurs by localized fracture and is typically a dynamic process. Recently, experiments on epithelial cell monolayers have revealed that this scenario can be significantly modified when the material susceptible to cracking is adhered to a hydrogel substrate. Thanks to the hydraulic coupling between the brittle layer and the poroelastic substrate, such a composite can develop a toughening mechanism that relies on the simultaneous growth of multiple cracks. Here, we study this remarkable behaviour by means of a detailed model, and explore how the material and loading parameters concur in determining the macroscopic toughness of the system. By extending a previous study, our results show that rapid loading conveys material toughness by promoting distributed cracking. Moreover, our theoretical findings may suggest innovative architectures of flaw-insensitive materials with higher toughness.  
\end{abstract}

\section{Introduction}

Nature has adopted diverse, remarkable strategies to enhance the flaw-tolerance of biological tissues, such as bone and tooth \cite{gao_materials_2003}. As a result, they can sustain relatively high levels of strain while maintaining their integrity. Epithelium is another example of such tissues. Despite the intrinsic brittle behavior of the cell monolayer \cite{Harris2012}, this can display a high fracture toughness, as it has recently been shown experimentally \cite{Casares2015}. In a previous work \cite{PhysRevLett.115.188105} we have demonstrated that this behavior is determined by the hydraulic coupling between the epithelial layer and the extracellular matrix, which can be regarded as a poroelastic, hydrogel-like material.
Specifically, because fracture requires flow into the crack to fill its volume, the kinetics of solvent migration within the hydrogel controls the velocity of crack propagation, so that decreased permeability promotes multiple-cracking at cell-cell junctions. 
Then, this distributed cracking mechanism maximizes the external work performed on the system before failure (which is a measure of toughness), since cell-cell separation requires a significant amount of work and is accompanied by dissipation due to solvent flow. 

The just described behavior sharply contrasts with that of brittle materials, which are highly flaw-sensitive and typically fail catastrophically, by localized fracture \cite{AndersonBook}. For instance, the equilibrium of two edge cracks with the same length in an elastic brittle layer under tensile load is unstable, so that any perturbation will cause only one of the cracks to propagate dynamically \cite{noselli_analysis_2013}. In general, since toughening of soft materials typically relies on energy dissipation \cite{Zhao2014}, studying the energetics of crack propagation \cite{Noselli2016} is crucial to understand the interplay of different physics involved in fracture phenomena and for the implementation of toughening strategies in synthetic materials \cite{Gong2014}. Therefore, we here reconsider the problem studied numerically in \cite{PhysRevLett.115.188105}, where a brittle layer containing two pre-cracks and bonded to a hydrogel substrate is subject to a remote strain.   In particular, we analyze the relative influence of the different dissipative mechanisms on the macroscopic toughness of the system. Moreover, we focus on the effect of the strain rate on the transition from distributed to localized cracking. 

The paper is organized as follows. We first establish the model accounting for the poroelasticity of the hydrogel substrate, the elasticity of the brittle layer and its hydraulic fracture. Then, the weak formulation of the model is presented, which is suitable for its implementation into a finite element code. Numerical results are shown and discussed concerning the dissipations for both the regimes of distributed and localized fracture. 

\section{Model for the hydraulic fracture of a brittle layer bonded to a hydrogel substrate}
\label{sec:computational}

In this section, we derive the governing equations for the nonlinear model of hydraulically driven crack propagation in a brittle layer bonded to a hydrogel substrate. Both the brittle solid and the hydrogel are modelled as layers of finite thickness and infinite length. To simplify notation, the derivation is presented for the case of a single crack. Then, in the following section, the numerical model is extended to account for two competing cracks. 

\subsection{Notation and kinematics}

\begin{figure}[b!]
\centering
\includegraphics[scale=1]{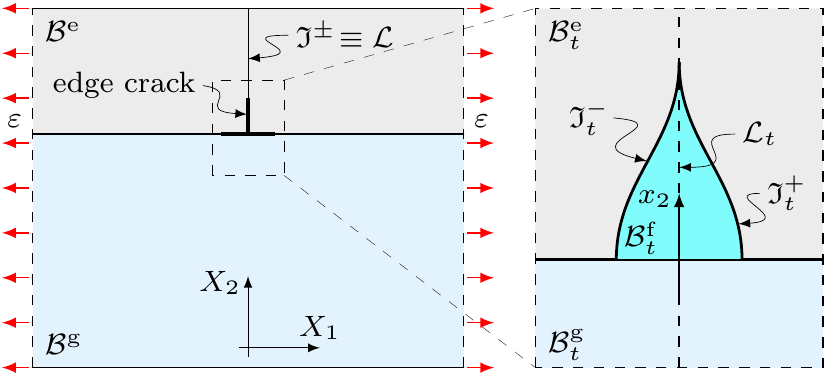}
\caption{Sketch of the reference configuration of the system (left) and of the current configuration of the edge crack (right) in the cut plane $X_3 = x_3 = 0$.}
\label{fig:configs}
\end{figure}

We introduce an orthonormal basis $\{\mb{e}_i\}$, $i=\{1,2,3\}$, for the three-dimensional Euclidean space $\mc{E}$. We denote by $\mc{B}^{\textrm{g}}_t \subset \mc{E}$ and $\mc{B}^{\textrm{e}}_t \subset \mc{E}$ the current configurations  at time $t$ of the hydrogel and the brittle layer, respectively; the corresponding reference configurations are indicated by dropping the subscript $t$. Further, we introduce two Cartesian coordinate systems $\{X_i\}$ and $\{x_i\}$, for the reference and the current configuration of the system, respectively. A material point $X$ of the reference solid domain $\mc{B} = \mc{B}^{\textrm{g}} \cup \mc{B}^{\textrm{e}}$ with coordinates $X_i$ is mapped into a spatial point $x$ with coordinates $x_i = f_i(X_j,t)$ in the current configuration $\mc{B}_t$  (Fig.~\ref{fig:configs}), where  $f_i$ are the coordinate representations of the motion of the solid domain. We will use the symbol $\mb{F}$ for the deformation gradient, and write $J=\det\mb{F}$ for its determinant and $\mb{F}^\star = J\mb{F}^{-T}$ for its cofactor. We assume plane strain conditions, such that all the quantities do not depend on $X_3$ and thus  $x_\alpha = f_\alpha(X_1,X_2,t)$, $\alpha = \{1,2\}$, whereas $ x_3 = f_3(X_3,t) = X_3$. We take all of the three-dimensional domains to be of unit-depth, so that, in the ensuing derivation, volume integrals and area integrals over the cross-section corresponding to the cut plane $X_3 = x_3 = 0$ coincide.

The brittle layer contains an initially closed pre-crack with length $a_{\rm{o}}$, whose faces belong to the referential segments $\mc{J}^\pm$, being $\mc{J}_t^-$ and $\mc{J}_t^+$ their current counterparts. For the reference configuration $\mc{L}$ of the longitudinal crack axis, we choose the segment aligned with the $X_2$-axis: $\mc{L}=\{ X \in \mc{E}\,|\,X_1 = X_3 = 0, X_2 \in [0,h]=\mc{H}\}$, where $h$ is the thickness of the brittle layer. We assume that the axis $\mc{L}_t$ remains straight and vertical upon deformation, and that the crack undergoes a plane motion symmetric with respect to such axis. Thus, the axis of the crack stretches by the amount $\lambda(X_2,t)  = \partial f^{\rm c}_2/\partial X_2(X_2,t)$ to follow the vertical motion $f_2^{\rm c}(X_2,t) =\lim_{X_1 \rightarrow 0^{\pm}}f_2(X_1,X_2,t)$, with $X_2 \in \mc{H}$.

We define the material description of the crack opening $\delta(X_2,t)$ as
\begin{align}
\delta(X_2,t) = \llbracket f_1(0,X_2,t) \rrbracket  = 2 f_1^{\rm c}(X_2,t)\,,
\end{align}
where the symbol $\llbracket \cdot \rrbracket$ denotes the jump operator, whereas $f_1^{\rm c}(X_2,t) = \lim_{X_1 \rightarrow 0^+} f_1(X_1,X_2,t)$ and the second equality follows from symmetry. By composing $\delta$ with the inverse of the deformation mapping, we obtain the crack opening as a function of the spatial location along $\mc{L}_t$: $\delta_{\rm s}(x_2,t) = \delta(X_2,t) \circ (f_2^{\rm c})^{-1}(x_2,t)$. The velocity of the crack faces may be readily computed as 
\begin{align}
\label{eq:vcrackfaces}
\mb{v}_{\rm c}^{\pm}(x_2,t) = \pm\frac{1}{2} \dot{\delta}_{\rm s}(x_2,t) \mb{e}_1 + w(x_2,t)\mb{e}_2\,,
\end{align}
where $w(x_2,t) = \dot{f}^{\rm c}_2(X_2,t) \circ (f_2^{\rm c})^{-1}(x_2,t)$ is its vertical component. Here, a superposed dot denotes the material time derivative (at $X_2$ fixed), so that
\begin{align}
\label{eq:matderiv}
\begin{split} 
\dot{\delta}_{\rm s}(x_2,t) &= \dot{\delta}(X_2,t) \circ (f_2^{\rm c})^{-1}(x_2,t) = \\[1.0mm]
&= \frac{\partial \delta_{\rm s}(x_2,t)}{\partial t} + w(x_2,t)\frac{\partial \delta_{\rm s}(x_2,t)}{\partial x_2}.
\end{split}
\end{align} 

We call $\mc{B}^{\rm{f}}_t \subset  \mc{E}$ the fluid domain, that is, the spatial volume with longitudinal axis $\mc{L}_t$ enclosed by the crack faces and filled with solvent. Finally, we denote by $\mb{v}(x,t),\ x\in\mc{B}_t^{\rm {f}}$, the spatial velocity field of the solvent within the crack. 

\subsection{Solvent flow within the crack}

In this section, we establish a reduced one-dimensional model for the solvent flow within the crack. Specifically, we derive a set of equations, defined over the axis $\mc{L}$, by localizing the integral balance laws for the fluid expressed in terms of resultant quantities over horizontal cross-sections. A related model was introduced in \cite{MOXReport} without accounting for stretching along the axis. We consider the steady flow of an incompressible, viscous fluid under the assumption that the motion is quasi-1D, \textit{i.e.} the velocity field $\mb{v}$ is independent of $x_3$ and is dominated by the longitudinal component $v_2 = \mb{v}\cdot\mb{e}_2$.

We define a control volume $\mc{V}_t \subset \mc{B}^{\textrm{f}}_t$, enclosed by the cross-sections that are located at two fixed positions $\bar{x}_2$ and $\bar{x}'_2$ along $\mc{L}_t$, and the crack faces $\mc{J}_t^{\pm}$.

\subsubsection{Conservation of solvent mass}

The equation of conservation of mass for the incompressible fluid in $\mc{V}_t$ reads
\begin{align}
\label{eq:incompressmassbal}
\int_{\partial\mc{V}_t}{\mb{v}\cdot\mb{n}\,\mbox{d}a}=0 \,,
\end{align}
where $\mb{n}$ is the outward unit normal to $\partial \mc{V}_t$. On the crack faces this is given by the following relation
\begin{align}
\label{eq:normcrackfaces}
\mb{n}_{\rm c}^{\pm} = \frac{1}{\| \mb{n}_{\rm c} \|} \left(\pm \mb{e}_1 - \frac{1}{2}\frac{\partial \delta_{\rm s}}{\partial x_2}\mb{e}_2\right) , 
\end{align}
with $\| \mb{n}_{\rm c} \| = \sqrt{1+1/4 (\partial \delta_{\rm s}/ \partial x_2)^2}$. Notice that the area element along the crack faces in Eq.~\eqref{eq:incompressmassbal} transforms according to $\mbox{d}a = \| \mb{n}_{\rm c} \|\mbox{d}x_2\mbox{d}x_3$. Using eqs.~\eqref{eq:vcrackfaces} and Eq.~\eqref{eq:normcrackfaces} the volume flux contributions over parts of $\mc{J}_t^{\pm}$ in Eq.~\eqref{eq:incompressmassbal} may be evaluated as
\begin{align}
\int_{\bar{x}_2}^{\bar{x}'_2}{\left(\dot{\delta}_{\rm s}-w\frac{\partial \delta_{\rm s}}{\partial x_2}\right)\mbox{d}x_2}\,, 
\end{align}
while the net flux through the cross-sections delimiting $\mc{V}_t$ is
\begin{align}
Q(\bar{x}'_2,t)-Q(\bar{x}_2,t)=\int_{\bar{x}_2}^{\bar{x}'_2}{\frac{\partial Q}{\partial x_2}\,\mbox{d}x_2}\,,
\end{align}
with
\begin{align}
Q(x_2,t) = \int_{-\delta_{\rm s}(x_2,t)/2}^{\delta_{\rm s}(x_2,t)/2}{v_2(x_1,x_2,t)\,\mbox{d}x_1}
\end{align}
the solvent volume flux per unit depth.
By summing these two contributions, Eq.~\eqref{eq:incompressmassbal} may be recast as
\begin{align}
\int_{\bar{x}_2}^{\bar{x}'_2}{\left(\dot{\delta}_{\rm s}-w\frac{\partial \delta_{\rm s}}{\partial x_2}+\frac{\partial Q}{\partial x_2}\right)\mbox{d}x_2}=0\,.
\end{align}
To express this equation in the reference configuration $\mc{L}$, we first observe that
\begin{align}
&\frac{\partial w(x_2,t)}{\partial x_2} = \left(\frac{1}{\lambda(X_2,t)}\frac{\partial \lambda(X_2,t)}{\partial t}\right) \circ (f_2^{\rm c})^{-1}(x_2,t) , \\[1.0mm]
&\frac{\partial \delta_{\rm s}(x_2,t)}{\partial x_2} = \left(\frac{1}{\lambda(X_2,t)}\frac{\partial \delta(X_2,t)}{\partial X_2}\right) \circ (f_2^{\rm c})^{-1}(x_2,t).
\end{align}
Then, upon exploiting Eq.~\eqref{eq:matderiv} and localizing, we obtain
\begin{align}
\label{eq:consmasscrack}
\frac{\partial\left(\delta\lambda\right)}{\partial t} + \frac{\partial q}{\partial X_2} = 0\,,
\end{align}
where $q(X_2,t) = q_{\rm s}(x_2,t) \circ f_2^{\rm c}(X_2,t)$ is the material description of the solvent flux
\begin{align}
q_{\rm s}(x_2,t) = Q(x_2,t)-w(x_2,t)\delta_{\rm s}(x_2,t)
\end{align}
relative to the material particles on the crack faces.

\subsubsection{Balance of forces}

Upon neglecting inertia, the balance of forces for the solvent within the crack along the longitudinal axis of $\mc{V}_t$ reads 
\begin{align}
\label{eq:intbalaforcesfluid}
\mb{e}_2\cdot\int_{\partial \mc{V}_t}{\mb{T}_{\rm f}\mb{n}\,\mbox{d}a}=0\,.
\end{align}
As a constitutive law for the stress tensor we take the representation for an incompressible, Newtonian viscous fluid, such that
\begin{align}
\mb{T}_{\rm f} = -p_{\rm s}\mb{I}+2\eta\,\rm{sym}(\grad{\mb{v}}) \,,
\end{align}
with $p_{\rm s}$ and $\eta$ the pressure and the viscosity of the solvent, respectively.
Following lubrication theory \cite{BatchelorBook}, we assume that: i) the velocity profile is parabolic along the cross-sections, \textit{i.e.}
\begin{align}
\label{eq:parabvel}
v_2(x_1,x_2,t)=4(w-v_2^{\rm max})\left(\frac{x_1}{\delta_{\rm s}}\right)^2+v_2^{\rm max} ,
\end{align}
where $v_2^{\rm max}=(3Q/\delta_{\rm s}-w)/2$;
ii) the pressure $p_{\rm s}(x_2,t)$ is uniform over each cross-section; iii) the components of the velocity gradient along $x_2$ are negligible. Under these hypotheses, the traction on the crack faces is
\begin{align}
\label{eq:fluidtraction}
\begin{split}
\mb{t}^\pm_{\rm f} &= \mb{T}_{\rm f}\mb{n}_{\rm c}^{\pm} \approx  \\[1.0mm]
&\approx -p_{\rm s}\mb{n}_c^{\pm}-6\eta \frac{q_{\rm s}}{\delta_{\rm s}^2}(\mb{e}_1\otimes\mb{e}_2+\mb{e}_2\otimes\mb{e}_1)\mb{n}_{\rm c}^{\pm}\,,
\end{split}
\end{align}
where in the last equality we have used the representation \eqref{eq:parabvel} evaluated at $x_2 = \pm\delta_{\rm s}/2$. With this, the contribution to the integral in \eqref{eq:intbalaforcesfluid} extending over the crack faces may be computed as
\begin{align}
\label{eq:pressint}
 &-12\int_{\bar{x}_2}^{\bar{x}'_2}{\frac{q_{\rm s}}{\delta_{\rm s}^2}\,\mbox{d}x_2} +\int_{\bar{x}_2}^{\bar{x}'_2}{p_{\rm s}\frac{\partial \delta_{\rm s}}{\partial x_2}\,\mbox{d}x_2}\,.
\end{align}
Moreover, the force resultant over the cross-sections at $\bar{x}_2$ and $\bar{x}'_2$ is
\begin{align}
\label{eq:viscstressint}
-p_{\rm s}(x_2,t)\delta_{\rm s}(x_2,t)\Big|^{\bar{x}'_2}_{\bar{x}_2} = -\int_{\bar{x}_2}^{\bar{x}'_2}{\frac{\partial (p_{\rm s}\delta_{\rm s})}{\partial x_2}\,\mbox{d}x_2} \,.
\end{align}
Finally, by expressing the sum of eqs.~\eqref{eq:pressint}-\eqref{eq:viscstressint} in the reference crack axis $\mc{L}$ and by localizing we obtain
\begin{align}
\label{eq:lubq}
q =- \frac{\delta^3}{12\eta\lambda}\frac{\partial p_{\rm f}}{\partial X_2}\,,
\end{align}
where $p_{\rm f}(X_2,t) = p_{\rm s}(x_2,t)\circ f_2^{\rm c}(X_2,t)$.

\subsection{Cohesive zone model}

To model crack propagation, we employ a cohesive zone approach \cite{FreundBook}. We take the  cohesive tractions $\mb{s}^{\pm}_{\rm c}$ per unit reference area  acting on the crack faces $\mc{J}^\pm$ to be orthogonal to the reference longitudinal axis $\mc{L}$ of the crack, and such that their magnitude is a function of the crack opening:
\begin{align}
\mb{s}^{\pm}_{\rm c} = \mp \sigma(\delta)\,\mb{e}_1\,.
\end{align}
We choose a bilinear traction-separation law for $\sigma(\delta)$, such that
\begin{align}
\sigma(\delta) = (1-d)\frac{\sigma_\textrm{o}}{\delta_\textrm{o}}\delta\,,
\end{align}
where $\sigma_\textrm{o}$ is the cohesive strength (attained for $\delta = \delta_\textrm{o}$) and $d(X_2,t)\in[0,1]$ is the damage. We prescribe the following evolution law for the damage:
\begin{align}
d = \frac{\delta_\textrm{c} (\delta^{\textrm{m}}-\delta_\textrm{o})}{\delta^{\textrm{m}}(\delta_\textrm{c}-\delta_\textrm{o})}\,,
\end{align}
where $\delta^{\textrm{m}}(X_2,t)$ is the maximum value attained by the opening during crack evolution and $\delta_\textrm{c}$ is the crack opening at failure ($d=1$). The fracture energy $\Gamma$ is related to the parameters of the cohesive zone model through the equation $\Gamma = \sigma_{\textrm{o}}\delta_\textrm{c}/2$.

In addition to the cohesive tractions, the crack faces $\mc{J}_t^{\pm}$ are subject to the pressure and to the shear stress exerted by the fluid, which contributes to the traction with the term $-\mb{t}_{\rm f}^{\pm}$, see Eq.~\eqref{eq:fluidtraction}.
The corresponding reference traction is
\begin{align}
\mb{s}^{\pm}_{\rm f} \!=\! \pm p_{\rm f}\mb{F}^\star\mb{e}_1\!+\! 6\eta\frac{q}{\delta^2}[(\mb{F}^{\star}\mb{e}_1\!\cdot\!\mb{e}_1)\mb{e}_2\!+\!(\mb{F}^{\star}\mb{e}_1\!\cdot\!\mb{e}_2)\mb{e}_1] ,
\end{align}
such that the total reference traction acting on the crack faces is given by
$\mb{s}^{\pm} = \mb{s}_{\rm f}^{\pm} + \mb{s}^{\pm}_{\rm c}$.

\subsection{Elasticity of the brittle layer}

By assuming inertia negligible, the balances of forces and moments in $\mc{B}^e$ read
\begin{align}
\label{eq:balaforces}
\divg \mb{S} = \mb{0}\,, && \mbox{skw}\,\mb{S}\mb{F}^{\rm T} = \mb{0}\,,
\end{align}
where $\mb{S}$ denotes the first Piola-Kirchhoff stress tensor. 

We model the brittle layer as an impermeable, elastic solid characterized by the compressible neo-Hookean free energy density
\begin{align}
\psi(\mb{F})=\frac{G_{\rm e}}{2}(\mb{F}\cdot\mb{F}-2\log J-3)+\frac{\Lambda}{2}(\log J)^2 ,
\end{align}
where $G_\textrm{e}$ and $\Lambda \gg G_\textrm{e}$ are the Lam\'e moduli. The corresponding first Piola-Kirchhoff stress is given by
\begin{align}
\label{eq:stressbl}
\mb{S} = \frac{\partial \psi}{\partial \mb{F}} =  G_{\textrm{e}}\left(\mb{F}-\frac{1}{J}\mb{F}^\star\right)+\Lambda\,\frac{\log J }{J}\mb{F}^\star\,.
\end{align} 
Notice that, because of Eq.~\eqref{eq:stressbl} and the plane strain hypothesis, $\mb{S}\mb{e}_1\cdot\mb{e}_3=\mb{S}\mb{e}_2\cdot\mb{e}_3=\mb{S}\mb{e}_3\cdot\mb{e}_1=\mb{S}\mb{e}_3\cdot\mb{e}_2=0$ and the out-of-plane balance equation in \eqref{eq:balaforces}$_1$ is trivially satisfied. Hence, the balance of forces and moments may be formulated in terms of the plane components of $\mb{S}$ only.

\subsection{Poroelasticity of the hydrogel layer}

In the following, we briefly summarize the non-linear swelling theory for hydrogels introduced in \cite{lucantonio_transient_2013}. The state of the hydrogel is described by the motion $f$ of the polymer network and the solvent concentration $c$ per unit reference volume. The chemical potential $\mu$ of the solvent within the hydrogel quantifies the energy carried by the solvent and represents the driving force of solvent migration. The corresponding solvent molar flux $\mb{h}$ characterizes the relative motion of the solvent with respect to the polymer matrix. Consistently with the plane strain hypothesis, we assume that solvent migration takes place in the plane $\mb{e}_1$-$\mb{e}_2$, so that $c(X_1,X_2,t)$, $\mu(X_1,X_2,t)$, and $\mb{h}\cdot\mb{e}_3 = 0$. 

The polymer matrix and the solvent are considered to be separately incompressible; hence, the change in volume of the hydrogel is related to the change in solvent concentration: 
\begin{align}
	\label{eq:swellconstr}
	J = 1 + \Omega (c-c_{\textrm{o}})\,,
\end{align}
where $\Omega$ is the solvent molar volume and $c_{\textrm{o}} = (J_{\textrm{o}}-1)/(\Omega J_{\textrm{o}})$ is the solvent concentration per unit reference volume associated to the initial free swelling from the dry configuration to $\mc{B}^{\textrm{g}}$. Here, $J_{\textrm{o}}$ is the volume ratio between the reference and the dry configuration. The constraint of Eq.~\eqref{eq:swellconstr} is enforced through the Lagrange multiplier $p$.

Swelling processes are governed by the equations of balance of forces and moments, which are the same as those for the brittle layer, see Eqs.~\eqref{eq:balaforces}, and by the balance of solvent mass in $\mc{B}^{\textrm{g}}$
\begin{align}
\label{eq:balasolgel}
\dot{c} = -\divg \mb{h}\,,
\end{align}
subject to the initial condition $c(X_1,X_2,0)=c_{\rm o}$.

As concerns the constitutive equations, we prescribe the following Flory-Rehner representation for the free energy density of the hydrogel \cite{doi_gel_2009,lucantonio_transient_2013}
\begin{align}
\psi(\mb{F},c) = \psi_{\rm e}(\mb{F}) + \psi_{\rm m}(c) \,,
\end{align}
where
\begin{align}
&\psi_{\rm e}(\mb{F})= \frac{G_{\rm d}}{2J_{\textrm{o}}}(J_{\textrm{o}}^{1/3}\mb{F}\cdot\mb{F}-3) 
\end{align}
and
\begin{align}
&\psi_{\rm m}(c)=\mc{R}Tc\left[\log\left(\frac{\Omega J_\textrm{o}c}{1+\Omega J_\textrm{o}c}\right)+\chi\frac{1}{1+\Omega J_\textrm{o}c}\right]
\end{align}
are the neo-Hookean elastic energy and the Flory-Huggins free energy of mixing, respectively. Here, $G_{\textrm{d}}$ is the shear modulus of the dry polymer, $\mc{R}$ is the universal gas constant, $T$ is the absolute temperature of the environment, and $\chi$ is the polymer-solvent mixing parameter.
For the consistency with thermodynamical principles, the corresponding constitutive equations are given by:
\begin{align}
&\mb{S} =  \frac{\partial \psi_{\rm e}}{\partial \mb{F}}-p\mb{F}^\star\,, \label{eq:stress} \\[1.0mm]
&\mu = \frac{\partial \psi_{\rm m}}{\partial c}+\Omega p\,, \label{eq:Lag} \\[1.0mm]
&\mb{h} = -\frac{c\,D}{\mc{R}T}\nabla\mu\,, \label{eq:solfluxconst}
\end{align}
where $D$ is the diffusivity of the solvent within the hydrogel. In the reference configuration $\mc{B}^{\textrm{g}}$, the hydrogel is in equilibrium with the external solvent whose chemical potential is $\mu_\textrm{o}$. Specifically, we consider an external solvent in equilibrium with its vapor, such that $\mu_\textrm{o} = 0\ \mbox{J/mol}$. Thus, the reference configuration is identified by the conditions $\mb{F}=\mb{I}$, $\mu = \mu_{\textrm{o}}$ and $\mb{S}=\mb{0}$, which together determine the swelling ratio $J_{\textrm{o}}$ as the solution of following equation
\begin{align}
\label{eq:freeswell}
\log{\left(1-\frac{1}{J_{\textrm{o}}}\right)} + \frac{1}{J_{\textrm{o}}} + \frac{\chi}{J_{\textrm{o}}^2} +\frac{G_{\textrm{d}}\Omega}{\mc{R}T} \frac{1}{J_{\textrm{o}}^{1/3}} = 0\,. 
\end{align}

\subsection{Boundary and interface conditions}
In the numerical implementation, we refer to a computational domain which is limited along the $\mb{e}_1$-axis by vertical boundaries of normal $\mb{m}=\pm \mb{e}_1$. In order for the computational model to appropriately approximate the theoretical setting, we choose its length to be sufficiently larger than the thickness of the composite.

Starting from the equilibrium state of free swelling with $\mu = \mu_{\rm o}$, the system is subject to a homogeneous, horizontal strain of constant rate $\dot{\varepsilon}$. Correspondingly, the vertical component $\mb{u}\cdot\mb{e}_1$ of the displacement field $\mb{u}(X_1,X_2,t)$ is prescribed along the vertical boundaries $\partial_{\mb u}\mc{B}$ of the composite, while the vertical component \mbox{$\mb{u}\cdot\mb{e}_2$} of the displacement field is constrained at the bottom boundary of the hydrogel only. In order to ease crack opening, continuity of the displacements is relaxed at the interface between the brittle layer and the hydrogel along small segments near the cracks. The remaining parts of the boundary $\partial\mc{B}$ are traction-free.

As regards the boundary conditions for solvent migration, we assume zero solvent flux on $\partial \mc{B}^{\textrm{g}}$. For short times, no significant exchange of solvent with the exterior can take place. At the crack inlet, we impose the continuity of solvent pressure: $\mu = \Omega p_{\rm f}$, where $\mu/\Omega$ is the pressure within the hydrogel. At the crack tip, we impose zero solvent flux: $q=0$.

\subsection{Power balance}

As mentioned in the introduction, we are interested in analyzing the breakdown of dissipative contributions upon the toughness enhancement of the system. To this aim, we write the total power balance for the brittle layer--hydrogel composite as
\begin{align}
&P_{\rm t} = \dot{E} + P_{\rm coh} + P_{\rm v} - P_{\rm s}
\end{align}
where
\begin{align}
&\dot{E} = \frac{\mbox{d}}{\mbox{d}t}\int_{\mc{B}}{\psi}\,, \\[1.0mm]
&P_{\rm t} = \int_{\partial_\mb{u}\mc{B}}{\mb{Sm}\cdot\dot{\mb{u}}}\,, \label{eq:Pt} \\[1.0mm]
&P_{\rm coh} = \int_{\mc{J}^+}{\sigma(\delta) \dot{\delta}} =  \int_{\mc{J}^-}{\sigma(\delta) \dot{\delta}}\,, \label{eq:Pcoh} \\[1.0mm]
&P_{\rm v} = 2\eta\int_{\mc{V}_t}{\|\mbox{sym}\nabla{\mb{v}}\|^2} \approx 12\eta\int_{\mc{L}}{\lambda \frac{q^2}{\delta^3}}\,, \\[1.0mm]
&P_{\rm s} =  -\int_{\mc{B}^{\rm g}}{\mb{h}\cdot\nabla \mu} \label{eq:Psol}
\end{align}
are the total free energy of the system, the mechanical power expended by the applied tractions, the power expended by the cohesive tractions in fracture processes, the power dissipated by transport of the fluid within the cracks, and  the power dissipated by solvent transport within the hydrogel, respectively. Notice that there is no contribution associated to solvent transport across the boundary of the system because we have assumed the hydrogel to be impermeable for short times. 

\subsection{Weak form of the governing equations}
In order to solve the problem set in the previous paragraphs by means of the finite element method, we recast the governing equations \eqref{eq:consmasscrack}, \eqref{eq:balaforces}$_1$, \eqref{eq:swellconstr}, and \eqref{eq:balasolgel}  in weak form. The continuity of the solvent pressure at the crack inlet (subscript `in') is enforced through the Lagrange multiplier $g$ corresponding to the (concentrated) solvent mass flux per unit length. Then, the weak formulation of the problem reads: find $p_{\rm f}$, $\mb{u}$, $c$, $p$ and $g$ such that the following equations
\begin{align}
&\int_{\mc{L}}{\left(-\frac{\partial}{\partial t}(\delta \lambda) \tilde{p}_{\rm f} + q \frac{\partial \tilde{p}_{\rm f}}{\partial X_2}\right)} + \Omega(g\,\tilde{p}_{\rm f})_{\rm in}= 0\,, \\[1.0mm]
&\int_{\mc{B}}{\mb{S}\cdot \nabla\tilde{\mb{u}}} = \int_{\mc{J}^+}{\mb{s}^+\cdot\tilde{\mb{u}}^+}+\int_{\mc{J}^-}{\mb{s}^-\cdot\tilde{\mb{u}}^-}\,, \\[1.0mm]
&\int_{\mc{B}^{\rm g}}{(J-1-\Omega c+\Omega c_{\rm o})\,\tilde{p}} = 0\,, \\[1.0mm]
&\int_{\mc{B}^{\rm g}}{(-\dot{c}\tilde{\mu}+\mb{h}\cdot \nabla\tilde{\mu})} - (g\,\tilde{\mu})_{\rm in} = 0\,, \\[1.0mm]
&(\mu(c,p)-\Omega p_{\rm f})_{\rm in}\,\tilde{g} = 0\,, 
\end{align}
hold for arbitrary test fields (indicated with a superposed tilde) compatible with the Dirichlet conditions. The weak form of the governing equations is complemented by Eq.~\eqref{eq:lubq} and by the constitutive relations \eqref{eq:stress}-\eqref{eq:solfluxconst}. This, and the corresponding boundary conditions, have been implemented into the finite element software COMSOL Multiphysics v5.2. Specifically, quadratic shape functions were used for all the unknown fields, except for the pressure field $p$, which was discretized using linear shape functions to get a reliable approximation of the volume constraint. The implicit, variable-order (from 1 to 5), adaptive step-size BDF solver was used for time-stepping. A quasi-Newton algorithm was employed to solve iteratively the non-linear algebraic system resulting from the finite element discretization at each time step. The direct solver MUMPS was chosen for the solution of the linearized system at each iteration. The mesh consisted of about $2 \times 10^4$ triangular elements corresponding to about $10^5$ degrees of freedom, and was made symmetric with respect to the vertical symmetry axis of the computational domain to avoid introducing any numerical bias in the distribution of solvent flux between the cracks. Local mesh refinement along the crack paths was performed to ensure that the cohesive zones were discretized with at least 20 elements. Notice that the faces $\mc{J}^+$ and $\mc{J}^-$ are modeled as distinct (but overlapping) segments in the numerical model where the tractions $\mb{s}^\pm$ are prescribed. 

\begin{figure*}[!h]	
	\centering
	\includegraphics[scale=1]{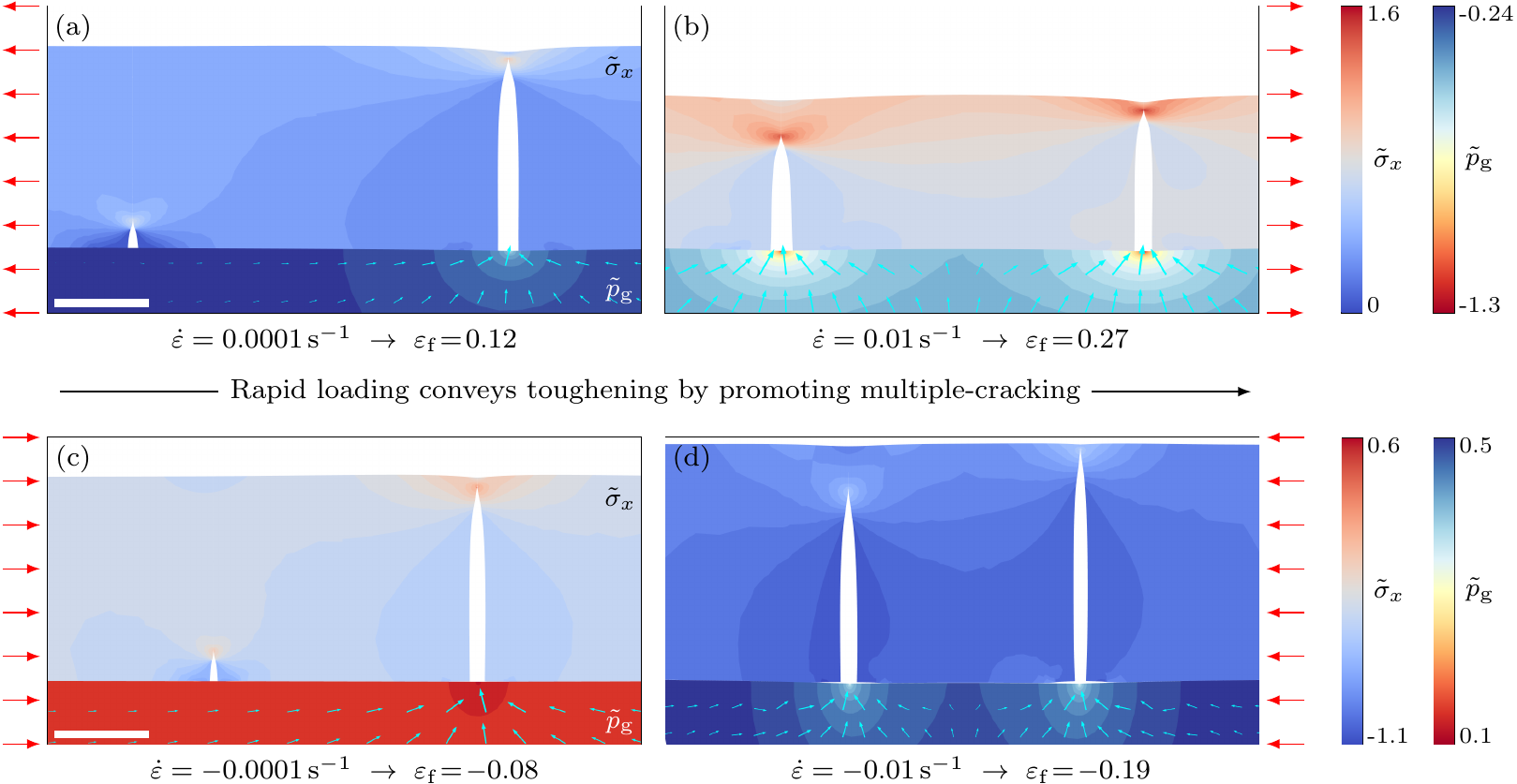}
	\caption{Snapshots of the system at failure as obtained from numerical simulations, for tensile (a),(b) and compressive (c),(d) strains. The transition from localized (a),(c) to distributed (b),(d) fracture corresponds to increasing values of the dimensionless group $\Pi = |\dot \varepsilon| s^2/D$, here obtained by varying the strain rate. Contours represent the longitudinal stress $\tilde{\sigma}_x = \mb{SF}^{\rm T}\mb{e}_{1}\cdot\mb{e}_1/JG_{\rm e}$ in the brittle layer and the solvent pressure $\tilde{p}_{\rm g} = \mu J^{1/3}_{\rm o}/(\Omega G_{\rm d})$ within the hydrogel, while $\varepsilon_{\rm f}$ is the value of the applied strain at failure. The scale bar is 5 $\mu$m, whereas the arrows in the hydrogel represent the solvent flux.}
	\label{fig:snapshots}
\end{figure*}

\begin{figure*}[h]	
	\centering
	\includegraphics[scale=1]{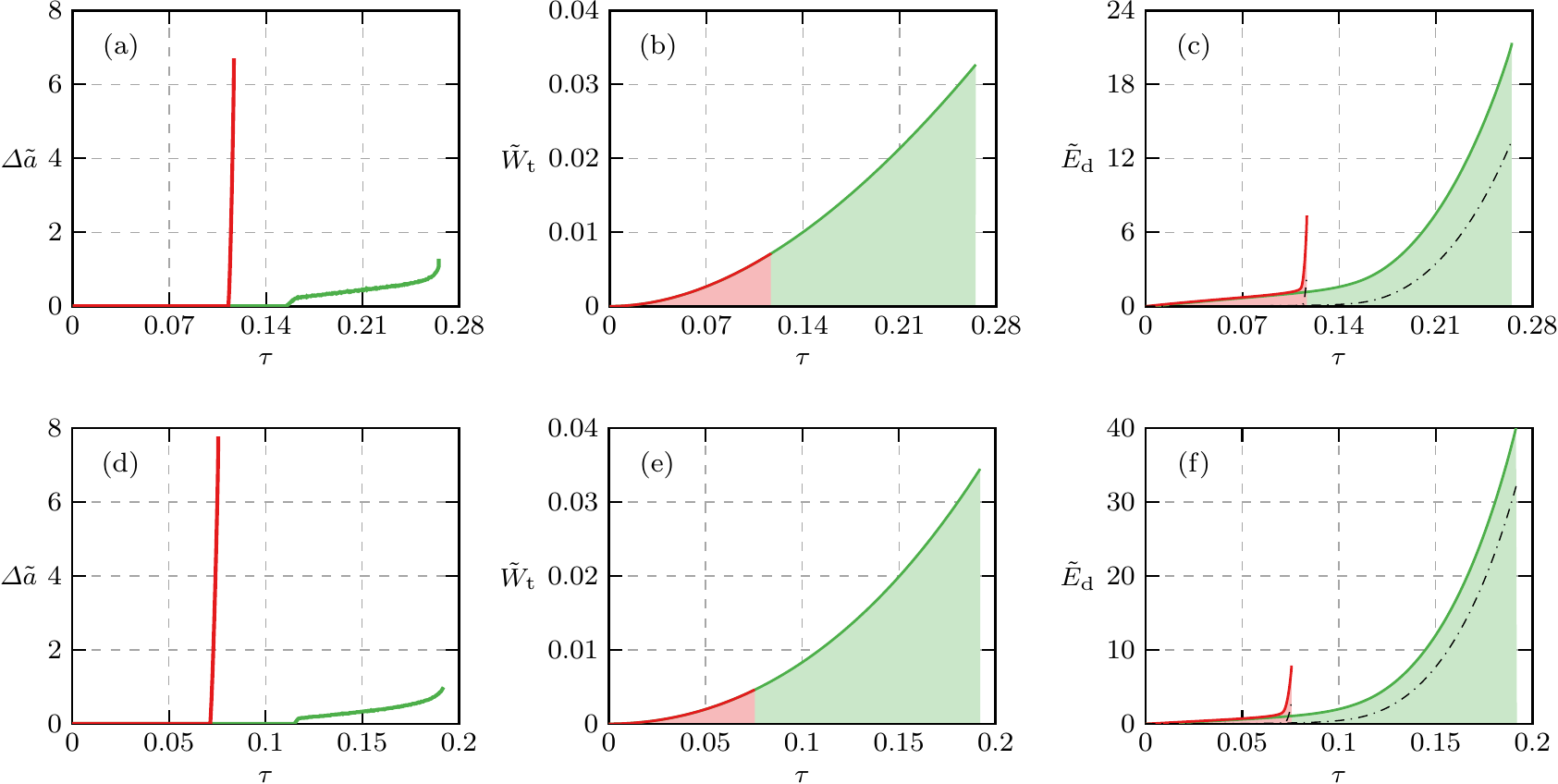}
	\caption{Dynamics and energetics of crack propagation in tension (a)-(c) and compression (d)-(f), for $|\dot{\varepsilon}|=0.0001\,\mbox{s}^{-1}$ (red, $\Pi = 2 \times 10^{-7}$) and $|\dot{\varepsilon}|=0.01\,\mbox{s}^{-1}$ (green, $\Pi = 2 \times 10^{-5}$). The plots show the relative crack length difference (a),(d), the external work performed on the system (b),(e) and the total dissipated energy (c),(f) as a function of time until failure. The dash-dotted line in (c),(f) represents the dissipated energy associated to solvent transport.}
	\label{fig:dissipations}
\end{figure*}

\section{Numerical results}

We apply the just described model to the analysis of the ideal case of a brittle layer containing two edge pre-cracks of initial length $a_{\rm o}$ and separation $s$. To break symmetry, we introduce a 10\% difference between the toughnesses $\Gamma_{\rm R}$ and $\Gamma_{\rm L}$ of the right and left crack, respectively.

\begin{table}[!b]
	\centering
	\caption{Values of the dimensionless groups used in the numerical simulations.}
	\label{tabparam1}
	\begin{tabularx}{\columnwidth/2}{Xc}
		\toprule
		\textbf{Parameter}							&	\textbf{Value} 	\\
		\midrule
		$h/a_{\rm o}$								&      $10$			\\
		$h/s$									&	$5/3$		\\
		$h/H$									&	$0.04$		\\
		$G_{\rm e}/\Lambda$						&	$0.02$ 		 \\
		$\Gamma_{\rm R}/\Gamma_{\rm L}$				&	$0.9$ 		 \\
		$\chi$									&	$0.46$		\\
		$\mu_{\rm o}/\mc{R}T$						&	$0$  			\\			
		$G_{\rm e} a_{\rm o}/\Gamma_{\rm L}$			&	$6.7$		 \\
		$\Gamma_{\rm L} a_{\rm o}/\eta D$				&	$2.5$		 \\
		\bottomrule
	\end{tabularx}
\end{table}

Dimensional analysis dictates that any relevant quantity, such as the total dissipated energy, has to depend on the following set of dimensionless groups: $h/a_{\rm o}$, $h/s$, $h/H$, $G_{\rm e}/\Lambda$, $G_{\rm d}/J^{1/3}_{\rm o} G_{\rm e}$, $G_{\rm d} \Omega/\mc{R}T$, $\Gamma_{\rm R}/\Gamma_{\rm L}$, $\chi$, $\mu_{\rm o}/\mc{R}T$, $|\dot \varepsilon| s^2/D$, $\tau$, $G_{\rm e} a_{\rm o}/\Gamma_{\rm L}$, $\Gamma_{\rm L} a_{\rm o}/\eta D$.
Here, $H$ is the thickness of the hydrogel substrate, whereas $\tau = |\dot\varepsilon| t$ is a dimensionless measure of time.

For the computational study, we set the values of part of these dimensionless groups as reported in Table~\ref{tabparam1}. The first three are geometrical ratios, while the following three are related to the elastic moduli of the brittle layer and the hydrogel. We recall from \cite{PhysRevLett.115.188105} that crack propagation under compression (tension) requires the ratio $3G_{\rm d}/4J^{1/3}_{\rm o}G_{\rm e}$ between the moduli of the hydrogel and the brittle layer to be higher (lower) than $1/2$. In compression, for instance, this condition allows the solvent pressure within the cracks to overcome the compressive stresses due to the remotely applied strain. Thus, we set $G_{\rm d}/J^{1/3}_{\rm o}G_{\rm e} = 0.38$, $G_{\rm d}\Omega/\mc{R}T = 6\times10^{-5}$ to study cracking in tension, and $G_{\rm d}/J^{1/3}_{\rm o}G_{\rm e} = 3.82$, $G_{\rm d}\Omega/\mc{R}T = 4\times10^{-4}$ for the simulations of cracking in compression.
Given $\chi$, $G_{\rm d}\Omega/\mc{R}T$ and $\mu_{\rm o}/\mc{R}T$, the initial swelling ratio $J_{\textrm{o}}$ is computed from Eq.~\eqref{eq:freeswell}. Eventually, by allowing the remaining parameters to vary, we regard a dimensionless, physical quantity as a function of $\Pi = |\dot\varepsilon| s^2/D$ and $\tau$.

We now focus on the significance of the dimensionless group of $\Pi$ upon the fracture behavior of the system. In a previous study \cite{PhysRevLett.115.188105}, we have demonstrated that decreased diffusivity conveys toughening by promoting multiple-cracking. Here, in light of dimensional analysis, we deduce that the transition from localized to distributed fracture is actually controlled by the dimensionless group $\Pi$. The results in Fig.~\ref{fig:snapshots}, where we report snapshots of the system at failure, confirm that such a transition is associated to an \textit{increase} in the value of $\Pi$, both in tension and compression. Further evidence is provided in Figs.~\ref{fig:dissipations}(a),(d), where the relative length difference $\Delta \tilde{a} = (a_{\rm R}-a_{\rm L})/a_{\rm o}$ between the cracks sharply grows for $\Pi = 2\times 10^{-7}$, as rapid, single crack propagation occurs. Therefore, multiple-cracking may be obtained by either an increase in strain rate or separation between the cracks, or by a decrease in diffusivity of the solvent within the hydrogel. 

The just described behavior may be rationalized as follows. In a brittle solid, the mechanism of crack tip shielding is responsible for localized, dynamic fracture. Any offset between the crack tips implies a progressive decrease (increase) in the energy release rate of the lagging (leading) crack \cite{noselli_analysis_2013}. In the presence of the hydrogel, crack advance always requires solvent transport and is thereby controlled by its diffusivity. Hence, continued loading that is rapid with respect to the kinetics of solvent transport can favour distributed cracking by sustaining the driving force of both fractures.  

Multiple-cracking results in an enhancement of the macroscopic toughness of the system. To quantify such an enhancement, we report in Figs.~\ref{fig:dissipations}(b),(e) the dimensionless mechanical work $\tilde{W}_{\rm t} = W_{\rm t}/(G_{\rm e}h+G_{\rm d}H)$ performed on the system as computed by time-integration of Eq.~\eqref{eq:Pt}. Notice that, an increase of two orders of magnitude in the strain rate produces more than a three-fold increase in the work at failure, both in tension and in compression. This trend is motivated by the increase of the dissipation $\tilde{E}_{\rm d}=E_{\rm d}/(\Gamma_{\rm L}+\Gamma_{\rm R})a_{\rm o}$ that accompanies distributed cracking, see Figs.~\ref{fig:dissipations}(c),(f). Here, $E_{\rm d}$ collects the time-integrals of the powers dissipated by fracture, Eq.~\eqref{eq:Pcoh}, and solvent transport, Eq.~\eqref{eq:Psol}, as the energy dissipated by viscous flow within the cracks is comparably negligible. Interestingly, we notice that dissipation due to solvent transport mainly contributes to such an increase in $\tilde{E}_{\rm d}$. As a consequence, the relative contribution of the work of fracture to $\tilde{E}_{\rm d}$ decreases in the transition from localized to distributed cracking. 

\section{Conclusions}

Motivated by recent experimental results on the fracture of epithelial cell monolayers adhered to an hydrogel substrate, a model has been developed that allows for the analysis of hydraulically driven cracking of this system. We have shown that such a composite can develop a toughening mechanism that relies on the multiple-cracking of the brittle phase. In particular, we have demonstrated that the transition from localized to distributed cracking is determined by either a decrease in solvent diffusivity or an increase in loading rate. A detailed energy analysis has quantified the dissipative processes associated with multiple-cracking. Future work will investigate upon the possibility of implementing this concept in the engineering of flaw-tolerant, biomimetic materials.

\section*{Acknoledgements}
The authors acknowledge the support of the European Research Council (AdG-340685 MicroMotility) and of National Group of Mathematical Physics (GNFM-INdAM) through the initiative ``Progetto Giovani''. The authors also thank prof. Robert M. McMeeking for useful discussions on the subject.


\begin{thebibliography}{10}
\providecommand{\url}[1]{{#1}}
\providecommand{\urlprefix}{URL }
\expandafter\ifx\csname urlstyle\endcsname\relax
\providecommand{\doi}[1]{DOI~\discretionary{}{}{}#1}\else
\providecommand{\doi}{DOI~\discretionary{}{}{}\begingroup
\urlstyle{rm}\Url}\fi

\bibitem{AndersonBook}
Anderson, T.L.: Fracture Mechanics: Fundamentals and Applications.
\newblock CRC Press (2005)

\bibitem{BatchelorBook}
Batchelor, G.K.: An Introduction to Fluid Dynamics.
\newblock Cambridge University Press, Cambridge (1967)

\bibitem{Casares2015}
Casares, L., Vincent, R., Zalvidea, D., Campillo, N., Navajas, D., Arroyo, M.,
  Trepat, X.: Hydraulic fracture during epithelial stretching.
\newblock Nature Materials \textbf{14}(3), 343--351 (2015).
\newblock \doi{10.1038/nmat4206}.
\newblock \urlprefix\url{http://dx.doi.org/10.1038/nmat4206}

\bibitem{doi_gel_2009}
Doi, M.: Gel dynamics.
\newblock J. Phys. Soc. Jpn. \textbf{78}(5), 052,001 (2009).
\newblock \doi{10.1143/JPSJ.78.052001}.
\newblock \urlprefix\url{http://journals.jps.jp/doi/abs/10.1143/JPSJ.78.052001}

\bibitem{MOXReport}
Formaggia, L., Veneziani, A.: Reduced and multiscale models for the human
  cardiovascular system.
\newblock Tech. Rep.~21, MOX-Politecnico di Milano (2003)

\bibitem{FreundBook}
Freund, L.B.: Dynamic Fracture Mechanics.
\newblock Cambridge University Press, Cambridge (1998)

\bibitem{gao_materials_2003}
Gao, H., Ji, B., J\"{a}ger, I.L., Arzt, E., Fratzl, P.: Materials become
  insensitive to flaws at nanoscale: {Lessons} from nature.
\newblock Proc. Natl. Acad. Sci. USA \textbf{100}(10), 5597--5600 (2003).
\newblock \doi{10.1073/pnas.0631609100}.
\newblock \urlprefix\url{http://www.pnas.org/content/100/10/5597}

\bibitem{Gong2014}
Gong, J.P.: Materials both tough and soft.
\newblock Science \textbf{344}(6180), 161--162 (2014).
\newblock \doi{10.1126/science.1252389}.
\newblock \urlprefix\url{http://science.sciencemag.org/content/344/6180/161}

\bibitem{Harris2012}
Harris, A.R., Peter, L., Bellis, J., Baum, B., Kabla, A.J., Charras, G.T.:
  Characterizing the mechanics of cultured cell monolayers.
\newblock Proc. Natl. Acad. Sci. USA \textbf{109}(41), 16,449--16,454 (2012).
\newblock \doi{10.1073/pnas.1213301109}.
\newblock \urlprefix\url{http://www.pnas.org/content/109/41/16449}

\bibitem{lucantonio_transient_2013}
Lucantonio, A., Nardinocchi, P., Teresi, L.: Transient analysis of
  swelling-induced large deformations in polymer gels.
\newblock Journal of the Mechanics and Physics of Solids \textbf{61}(1),
  205--218 (2013).
\newblock \doi{10.1016/j.jmps.2012.07.010}.
\newblock
  \urlprefix\url{http://linkinghub.elsevier.com/retrieve/pii/S0022509612001548}

\bibitem{PhysRevLett.115.188105}
Lucantonio, A., Noselli, G., Trepat, X., DeSimone, A., Arroyo, M.: Hydraulic
  fracture and toughening of a brittle layer bonded to a hydrogel.
\newblock Physical Review Letters \textbf{115}, 188,105 (2015).
\newblock \doi{10.1103/PhysRevLett.115.188105}.
\newblock
  \urlprefix\url{http://link.aps.org/doi/10.1103/PhysRevLett.115.188105}

\bibitem{noselli_analysis_2013}
Noselli, G., Deshpande, V.S., Fleck, N.A.: An analysis of competing toughening
  mechanisms in layered and particulate solids.
\newblock International Journal of Fracture \textbf{183}(2), 241--258 (2013).
\newblock \doi{10.1007/s10704-013-9890-8}.
\newblock
  \urlprefix\url{http://link.springer.com/article/10.1007/s10704-013-9890-8}

\bibitem{Noselli2016}
Noselli, G., Lucantonio, A., McMeeking, R.M., DeSimone, A.: Poroelastic toughening in polymer gels: A theoretical and numerical study.
\newblock Journal of the Mechanics and Physics of Solids \textbf{94}, 33--46
  (2016).
\newblock \doi{10.1016/j.jmps.2016.04.017}.
\newblock
  \urlprefix\url{http://www.sciencedirect.com/science/article/pii/S0022509616301818}

\bibitem{Zhao2014}
Zhao, X.: Multi-scale multi-mechanism design of tough hydrogels: building dissipation into stretchy networks.
\newblock Soft Matter \textbf{10}(5), 672--687 (2014).
\newblock \doi{10.1039/C3SM52272E}.
\newblock \urlprefix\url{http://dx.doi.org/10.1039/C3SM52272E}

\end{thebibliography}
\end{document}